\documentclass{article}
\usepackage{amsfonts}
\usepackage[margin=1.0 in]{geometry}
\usepackage{cite}

\newtheorem{theorem}{Theorem}
\newtheorem{acknowledgement}[theorem]{Acknowledgement}

\input{tcilatex}
\begin{document}

\title{Comments on the Weyl-Wigner calculus for lattice models }
\author{Felix A. Buot \\
Center for Theoretical Condensed Matter Physics (CTCMP), \\
Cebu Normal University, Cebu City 6000, Philippines \\
LCFMNN, Department of Physics, University of San Carlos, \\
Talamban, Cebu City 6000, Philippines\\
C\&LB Research Institute, Carmen, Cebu 6005, Philippines}
\date{}
\maketitle

\begin{abstract}
Here, we clarify the physical aspects between the discrete Weyl-Wigner (W-W)
formalism, well developed in condensed matter physics, and the so-called
'precise Weyl-Wigner calculus for lattice models' recently appearing in the
literature. We point out that the use of compact continuous momentum space
for a discrete lattice model is unphysically founded. It has an
incommensurate phase space, highly unphysical, lacks the finite fields
aspects, as exemplified by the Born-von Karman boundary condition of
compactified Bravais lattice of solid-state physics, and leads to several
ambiguities. This new W-W formalism simply lacks bijective Fourier
transformation, which is well-known to support the uncertainty principle of
canonical conjugate dynamical variables of quantum physics. Moreover, this
new W-W formalism for lattice models failed to handle the quantum physics of
qubits, representing two discrete lattice sites.
\end{abstract}

\section{Introduction}

There has been a surging interest in discrete Weyl-Wigner formulation of
quantum physics in recent years \cite{fial, liga, kasper}. Several recent
W-W formulations for discrete lattice models have appeared in the
literature. Here we give some clarifying comments on the physical merits of
these various formulations. Specifically, our comments is esentially focused
on the so-called precise W-W calculus for lattice models that has most
recently appeared in the literature \cite{fial}.

The discrete Weyl-Wigner (W-W) quantum physics has been well-developed in
condensed matter physics with its various successful applications, including
the IQHE \cite{buot1, buot2, buot3, buot4, buot5, buot6, buot7, buot8,
buot9, jb}. It seems that it just needs to be properly applied or adapted to
quantum field theory with ultraviolet cut-of. Below, I present a counter
example that exposes the weakness, unnecessary complications, and
corresponding physical and mathematical ambiguities of the new
\textquotedblleft precise\textquotedblright\ W-W formulation of discrete
lattice models that has recently appeared, as well as other discrete W-W
formulations, in the literature.

\section{Wannier functions and Bloch functions physics}

First of all, there are already l\textit{ong-standing} \textit{existing
quantum models in physics} that has firmly and physically guided the
discrete phase-space physics of the W-W formulation in condensed matter
physics namely, (a) \textit{Localized Wannier function} and \textit{extended
Bloch function} for discrete lattice in solid state physics, obeying the
Born-von Karman boundary condition (or strictly speaking, \textit{modular
arithmetic based on finite fields, }akin to a group theory of integers), (b)
Dirac delta function and plane waves in the continuum limit, i.e., phyically
and more importantly, \textit{only for continuous coordinate space can one
have continuous momentum space} (this quantum mechanical principle is
violated in a recent unphysically-formulated W-W formalism, the so-called
'precise' Weyl-Wigner calculus for lattice models, where \textit{discrete
lattice} coordinates is unphysically and incommensurately matched with 
\textit{compact continuous} momentum space)

In both (a) and (b), we have the eigenvector for positions (or discrete
lattice position), $\left\vert q\right\rangle $, and eigenvector for
momentum (or discrete crystal momentum), $\left\vert p\right\rangle $. Of
course, all respective eigenspaces only go to continuum spaces in the limit
of lattice constant goes to zero as in (b).

These respective eigenvectors, $\left\vert q\right\rangle $ and $\left\vert
p\right\rangle $, in (a) and (b) are related bijectively by Fourier
transformation, $\left\vert q\right\rangle $ to $\left\vert p\right\rangle $
via $\left\langle p\right. \left\vert q\right\rangle ,$ and often produces
results \textit{akin} to quantum uncertainty principle in their
probabilistic coordinate components. Thus, to construct a physically-based
discrete phase space or discrete W-W quantum physics one must be guided by
the following observations.

\section{Construction of W-W formalism: Discrete phase-space based on finite
fields}

The physically-based construction of W-W formalism in condensed-matter
physics is guided by the following well-known aspects of solid-state physics 
\cite{buot1}. (i) The invariance of this physical scheme of \textit{complete
and orthogonal} set of \ $\left\{ \left\vert q\right\rangle \right\} $ and 
\textit{complete and orthogonal set} of $\left\{ \left\vert p\right\rangle
\right\} ,$ in going from discrete to continuum physics, provides a \textit{%
strong guide} for formulating discrete quantum phase space W-W formalism in
condensed matter physics, (ii) Another \textit{crucial guidance }comes from
the observation that the the number of discrete lattice points (and hence
the number of discrete momentum points) must be an \textit{odd} prime number
for obvious inversion symmetry reason. Moreover, all arithmetic oprerations
on this group of numbers must be closed, i.e., all arithmetic operation must
be a modular arithmetic with \textit{prime number modulus} (\textit{akin to
a group operation on prime number of integers}). In short, all arithmetic
operation on these numbers is a modular arithmetic based on finite fields,
since only for finite fields with prime number modulus does every nonzero
element have well-defined multiplicative inverse, and hence modular division
operation also provide closure.

\section{Generalization to other discrete quantum and classical systems}

\textit{First Generalization:} Guidance (i) and (ii) allow us to generalize
discrete phase space based on finite fields to be particularly useful when
the quantum numbers involved, specifying the quantum states, are discrete
configurations other than the particle position and momentum quantum numbers 
\cite{buot8}. A simplest example is that of quantum bit or qubit, which will
be discussed below. \textit{Second Generalization:} The crucial importance
and power of using finite fields is that one can easily generalized the
discrete Wigner distribution construction based on the algebraic concept of
finite fields, which are extension of prime fields, where $q$ and $p$ are
field elements ($mod$ \textit{irreducible polynomial}), useful in quantum
computing, visualization, and communication/information sciences. Here we
have $p^{n}$ elements for some prime $p$ and some integer $n$ $>1$ useful
for constructing the Wigner distribution function for spin=$\frac{1}{2}$%
systems \cite{gibbons, buot8}.

I will detail below to show that the new 'precise' W-W formulation of
lattice models is not consistent with the above-mentioned physically based
models. It is essentially the misguided use of discrete lattice positions
coupled with the \textit{compact} \textit{continuous} momentum space that
complicate the new 'precise' W-W formulation and renders several ambiguities
by incurring a non-bijective canonical conjugate dynamical variables,
momentum $p$ and coordinate $q$. This, and \textit{together with the lack of
modular arithmetic on finite fields}, which is not invoked at all in this
new W-W formulation is adding to several more ambiguities. These ambiguities
stemmed from the use of the \textit{unphysical compact continuous momentum
space corresponding to a discrete lattice space models}.

\section{The W-W formulation in condensed-matter physics}

Let me first summarize the Bravais lattice vectors and their corresponding
reciprocal lattice vectors, since this points to some generalities of the
discrete W-W formulation in condensed matter physics. In $2$-D lattice, we
have the reciprocal lattice vectors, $\vec{b}_{1}$ and $\vec{b}_{2}$ given
by the matrix,%
\begin{equation}
\left( 
\begin{array}{cc}
\vec{b}_{1} & \vec{b}_{2}%
\end{array}%
\right) =\frac{1}{\hat{n}\cdot \left( \vec{a}_{1}\times \vec{a}_{2}\right) }%
\left( 
\begin{array}{cc}
\vec{a}_{2}\times \hat{n} & \hat{n}\times \vec{a}_{1}%
\end{array}%
\right)  \label{eq. 2-4}
\end{equation}%
which geometrically means that $\vec{b}_{1}$ is perpendicular to $\vec{a}%
_{2} $ and $\vec{b}_{2}$ perpendicular to $\vec{a}_{1}$. The $\hat{n}$ is
the unit vector normal to the $2$-D lattice plane.

In $3$-D lattice, the reciprocal lattice vectors, $\vec{b}_{1}$, $\vec{b}%
_{2} $, and $\vec{b}_{3}$ are contained in the matrix%
\begin{equation}
\left( 
\begin{array}{ccc}
\vec{b}_{1} & \vec{b}_{2} & \vec{b}_{3}%
\end{array}%
\right) =\left( 
\begin{array}{ccc}
\frac{\vec{a}_{2}\times \vec{a}_{3}}{\vec{a}_{1}\cdot \left( \vec{a}%
_{2}\times \vec{a}_{3}\right) } & \frac{\vec{a}_{3}\times \vec{a}_{1}}{%
\left( \vec{a}_{3}\times \vec{a}_{1}\right) \cdot \vec{a}_{2}} & \frac{\vec{a%
}_{1}\times \vec{a}_{2}}{\left( \vec{a}_{1}\times \vec{a}_{2}\right) \cdot 
\vec{a}_{3}}%
\end{array}%
\right)  \label{eq 2-5}
\end{equation}%
This is the form usually given in solid-state physics textbooks. It is worth
mentioning that the above formulas are independent of any choosen coordinate
system. Note that the matrix multiplication given by 
\begin{equation}
\left( M\right) \left( M\right) ^{-1}=I  \label{eq. 2-6}
\end{equation}%
becomes 
\begin{eqnarray}
\left( 
\begin{array}{ccc}
\vec{a}_{1} & \vec{a}_{2} & \vec{a}_{3}%
\end{array}%
\right) \left( 
\begin{array}{ccc}
\vec{b}_{1} & \vec{b}_{2} & \vec{b}_{3}%
\end{array}%
\right) ^{T} &=&I  \nonumber \\
\left( 
\begin{array}{ccc}
\vec{a}_{1} & \vec{a}_{2} & \vec{a}_{3}%
\end{array}%
\right) _{\text{vectors in column}}\left( 
\begin{array}{c}
\vec{b}_{1} \\ 
\vec{b}_{2} \\ 
\vec{b}_{3}%
\end{array}%
\right) _{\text{vectors in rows (adjoint)}} &=&I_{3\times 3}  \label{eq. 2-7}
\end{eqnarray}

\subsection{Translation symmetry}

The overarching concept in solid state physics is the concept of translation
symmetry along any symmetry directions of the lattice. Inversion symmetry of
the lattice structure is also one of the symmetry properties. Thus, the Buot
discrete phase space W-W formalism in condensed matter physics is compatible
with any of the lattice structures defined by Eq. (\ref{eq. 2-4}) -- (\ref%
{eq 2-5}), i.e., not limited to cubic lattice structures only.

\subsubsection{Discrete momentum space}

Each energy band corresponds to the splitting of the energy levels of one
atomic site into $N$ levels where $N$ is the number of lattice sites in a 
\textit{compactified} Bravais lattice obeying the Born-von Karman boundary
condition in a given symmetry direction. This is the basis of energy band
quantum dynamics. Therefore, the number of crystal momentum states in each
band (in Brillouin zone) is exactly equal to the number of lattice sites.
This is an important physical observation which is violated by the new
'precise' W-W calculus of lattice models. Hence, there has to be a bijective
mapping between number of discrete lattice sites and the number of discrete
crystal momentum states. This is the essence of the powerful theoretical
concept of localized function around each lattice site, the so-called 
\textit{Wannier function }$\left\vert q\right\rangle $, and the extended
function over all lattice points, the so-called \textit{Bloch function }$%
\left\vert p\right\rangle $, related through the bijective discrete Fourier
transformation. The bijectivity aspect is crucial here to avoid ambiguities.

\subsubsection{Bijective Fourier transformation}

In short, the crystal momentum space is essentially discrete to yield a
bijective mapping to the discrete lattice sites in a discrete Fourier
transformation, it should definitely not be a compact continuous momentum
space, which is the highly unphysical basic assumption of the 'precise' W-W
formulation. This would give an ambiguous or ill-defined correspondence
between lattice sites and continuous momentum space in a Fourier
transformation, a bijective deficiency. Moreover, without the provision of
closure property based on modular arithmetic of the mathematics of finite
fields, the use of continous momentum space will lead to other ambiguities
in taking the Weyl transform. To cure this deficiency would only lead to
obscure and complex mathematics far from the simple basic physics, which is
the characteristic of the new 'precise' W-W formalism, and other recent
discrete W-W formalisms in the literature.

The Bloch function is an eigenfunction of crystal momentum and the Wannier
function is the eigenfunction of lattice site position. Observe that these
eigenfunctions of phase space operators are well established for gapped
structures with translational symmetry, or energy band far removed from the
other energy bands. Generalized Wannier function can also be defined for
coupled energy bands, using decoupling scheme like the Foldy-Woutheysen
transformation for relativistic Dirac electrons. Indeed, it has been shown
that counterparts of Wanner function and Bloch function exist for the
decoupled positive energy states of relativistic electrons, with the '%
\textit{Dirac}-\textit{Wannier function}' localization about the size of
Compton wavelength. Electric Wannier function and magnetic Wannier function
also exist, as well as their respective Bloch functions, for uniform
external electromagnetic fields. This physical idea has been extended by
Buot to formally construct the discrete phase space quantum mechanics 
\textit{based on finite fields} for cases where $q$ and $p$ are not position
and momentum variables, useful for quantum computing.

Now for the finite number of lattice points to have \textit{closure property}
under modular arithmetic operations of addition (includes inversion),
multiplication, and division, the number of lattice points must be a prime
number obeying the Born-von Karman boundary condition, i.e., of prime
modulus obeying modular arithemetic of finite fields.

In general, the Buot formalism for discrete phase space in condensed matter
physics is based on the mathematics of finite fields \cite{gibbons} and
holds for any prime number\footnote{%
Although every finite field, with $p^{n}$ elements for some prime $p$ and
some integer $n$ $\succeq 1$, often deals with irreducible polynomials over
ring $%
\mathbb{Z}
$ of integers, or over\ field $%
\mathbb{Q}
$ of rational numbers, or over field $%
\mathbb{R}
$ of real numbers, or over field $%
\mathbb{C}
$ of complex numbers, the role of irreducible polynomials can be played by
prime numbers themselves for $n=1$: prime numbers (together with the
corresponding negative numbers of equal modulus) are the irreducible
integers. They exhibit many of the general properties of the concept
'irreducibility' that equally apply to irreducible polynomials, such as the
essentially unique factorization into prime or irreducible factors: Every
polynomial $p(x)$ in ring of polynomials with coefficients in $F$, denoted
by $F\left[ x\right] $, can be factorized into polynomials that are
irreducible over $F$. This factorization is unique up to permutation of the
factors and the multiplication of constants from $F$ to the factors.
\par
The simplest case of interest in Buot discrete W-W formulation is when $n=1$%
. In this case the finite field $GF\left( p\right) $ is the ring $\frac{Z}{pZ%
}$. This is a finite field with $p$ elements, usually labelled $0,1,2,...p-1$%
, where arithmetic is performed modulo $p$, where nonzero elements have
multiplicative inverses$.$} of lattice points obeying the Born-von Karman
boundary condition. i.e., of prime modulus. It even holds for the most
elementary prime number $2$ of lattice points, to yield the $2\times 2$
Pauli spin matrices and the well-known \textit{Hadamard transformation}
between \textquotedblleft Wannier function\textquotedblright\ and
\textquotedblleft Bloch function\textquotedblright , i.e., discrete Fourier
transformation of two points in \textquotedblleft phase
space\textquotedblright\ leading to \textit{transformation of qubits}. Here
the \textquotedblleft Wannier function\textquotedblright\ and
\textquotedblleft Bloch function\textquotedblright\ simply become a guiding
theoretical concept, i.e, has acquired the status of a simple theoretical
device for discrete quantum physics. Indeed, the Buot formalism also gives
the generalized Pauli spin operators for any given prime number of lattice
points. It has also yielded all the entangled basis states for two and three
qubits, crucial to the physics of quantum teleportation \cite{buot8}.

\subsubsection{Phase-space point projectors in Hilbert space}

In Buot W-W formalism, any quantum mechanical operator is meaningfully
expanded in terms of \textit{phase-space point }projectors\textit{\ }in%
\textit{\ }Hilbert space. The \textit{coefficient of expansion} is precisely
the phase-space distribution function (lattice Weyl transform), or the
Wigner distribution function if the density operator is the one expanded in
terms of \textit{phase-space point }projectors \cite{buot8}. The same
phase-space point projector was obtained by Gibbons et al \cite{gibbons}
constructed from the eigenstate of the equation of lines in discrete $\left(
p,q\right) $-phase space based on finite fields.

\section{Violations and weakness of the new 'precise' W-W formulation}

So, we see that the use of the theoretical and physically-based model
pioneered by the concept of Wannier function and Bloch function in
solid-state physics is essential to a \textit{formal discretization in phase
space quantum physics based on the mathematics of finite fields}. In
contrast, the basic assumption of the new 'precise' W-W calculus for lattice
models \cite{fial} has the following \textit{unphysical ingredients}, some
listed as follows.

$\left( 1\right) $ It misses the use of Hilbert space of discrete lattice
position eigenvectors and \textit{discrete} crystal momentum eigenvectors, a
powerful theoretical device for energy-band gapped structures,

$\left( 2\right) $ It has a complete lack of the operational concept of
modular arithmetic on finite fields, modulus prime number essential for
modular-arithmetic closure property of a finite prime number of lattice
points, \textit{to avoid ambiguous arithmetic manipulations} [continuum
limit only when lattice constant goes to zero],

$\left( 3\right) $ It has incurred an unnecessary complications by
considering two discrete lattices, \textit{physical} and ancillary \textit{%
unphysical} lattice, and the momentum space is erroneously (unphysically)
assumed as a compact continuous space, this coupled with being devorced from
the modular arithmetic of finite fields makes a direct Weyl transformation
on momentum space totally ambiguous and not bijective.

$\left( 4\right) $ It has violated the bijective mapping in a Fourier
transformation of discrete lattice position to discrete momentum phase
space, rendering this ambiguous, i.e., not bijective since the momentum
space is assumed compact continuous, incommensurate with the number of
discrete lattice sites, and

$\left( 5\right) $ It needs a correct physically-based assumption, namely,
that both spaces, lattice sites and momentum states, are discrete and
exactly equal in number of points, i.e., appropriately a prime number
obeying modular arithmetic of finite fields.

\subsection{Counter example}

A counter example can easily be given in which the use of discrete `lattice
points' paired with compact continuous momentum points \cite{fial} will fail
and does not make any sense at all. Take the simplest prime number $2$ of
lattice points $\left[ 0\text{ \ and\ }1\right] $, a qubit. If we follow the
new `precise' W-W formulation, the momentum space is a \textit{compact
continuous} space in the interval $\left\{ 0,\ \pi \right\} $ assuming
lattice constant of unity. Clearly the resulting Fourier transformation
between the two spaces is highly ambiguous and is not bijective at the very
least. On the other hand, if we follow Buot W-W formulation in condensed
matter physics based on the mathematics of finite fields of the prime
modulus number $2$, and the use of theoretical device in terms of Wannier
function and Bloch function, a lot of interesting physics is revealed \cite%
{buot8}. Indeed, the $2\times 2$ Pauli spin matrices emerge, as well as the
well-known Hadamard transformation, which has become the standard
transformation of a quantum bit or \textit{qubit} in quantum computing.

The Hadamard matrix is the transformation from the "\textit{Wannier function}%
", $\left\vert q\right\rangle $, to the corresponding "\textit{Bloch function%
}", $\left\vert p\right\rangle $, for the two-state system. Consider the
identity 
\[
\left\vert p\right\rangle =\dsum\limits_{q}\left\langle q\right\vert \left.
p\right\rangle \ \left\vert q\right\rangle , 
\]%
where the $\left\langle q\right\vert \left. p\right\rangle $ is the
transformation function. For discrete quantum mechanics, this is given by
the discrete Fourier transform function, 
\[
\left\langle q\right\vert \left. p\right\rangle =\frac{1}{\sqrt{N}}\exp
\left( -\frac{i}{\hbar }p\cdot q\right) . 
\]%
Upon substituting the possible values of $q$ and $p$, namely, $\left[ 0\text{
\ and\ }1\right] $ and $\left[ \frac{2\pi \hbar 0}{2}\text{ \ and\ }\frac{%
2\pi \hbar 1}{2}\right] $, respectively, for our two-state system, we
obtained the matrix for $\left\langle q\right\vert \left. p\right\rangle $%
\[
\left\langle q\right\vert \left. p\right\rangle =\frac{1}{\sqrt{2}}\left( 
\begin{array}{cc}
1 & 1 \\ 
1 & -1%
\end{array}%
\right) \equiv H, 
\]%
which is the Hadamard matrix, $H$, which is really a "two-state discrete
Fourier transform matrix".

\section{Concluding Remarks}

In summary, the method of \ new `precise' W-W calculus for lattice models 
\cite{fial} seems unnecessarily complicated, is not physically founded, and
is beset with ambiguities cited above. More importantly, it fails to handle
the physics of the simplest prime number $2$ of discrete lattice sites. At
the very least the cited ambiguities and bijective deficiency of the
so-called new precise Weyl-Wigner calculus for lattice models renders this
instead as an 'imprecise' W-W formulation for lattice models. It is worth
mentioning that the Buot W-W discrete phase space formalism has also been
the basis of numerical Monte Carlo approach to quantum transport \cite%
{salvino, rossi}.

\begin{acknowledgement}
The author is grateful to Allan Roy Elnar, Gibson Maglasang, and Dr. Roland
E.S. Otadoy for interesting discussions. He is grateful to the PCIEERD-DOST,
Philippines, for the Visiting Professor support at the Cebu Normal
University.
\end{acknowledgement}

\end{document}